\documentclass[aps,pra,twocolumn,superscriptaddress,showpacs,floatfix]{revtex4-1}
\usepackage[version=3]{mhchem}
\usepackage[english]{babel}
\usepackage{dcolumn}
\usepackage{graphicx}
\usepackage{amsmath}
\usepackage{longtable}
\usepackage{bigstrut}
\usepackage{chemfig}
\usepackage{tabularx}
\usepackage{multirow}
\usepackage[colorlinks=true, urlcolor=blue, citecolor=blue, filecolor=blue, linkcolor=blue]{hyperref}
\allowdisplaybreaks[1]

\begin{document}

\title{Deceleration and trapping of ammonia molecules in a traveling wave decelerator}

\author{Paul Jansen}
\affiliation{LaserLaB, Department of Physics and Astronomy, VU University Amsterdam, 
De Boelelaan 1081, 1081 HV Amsterdam, The Netherlands}
\author{Marina Quintero-P\'{e}rez}
\affiliation{LaserLaB, Department of Physics and Astronomy, VU University Amsterdam, 
De Boelelaan 1081, 1081 HV Amsterdam, The Netherlands}
\author{Thomas E. Wall}
\affiliation{LaserLaB, Department of Physics and Astronomy, VU University Amsterdam, 
De Boelelaan 1081, 1081 HV Amsterdam, The Netherlands}
\author{Joost~E.~van~den~Berg}
\affiliation{University of Groningen, KVI, Zernikelaan 25, 9747 AA, Groningen, The Netherlands}
\author{Steven Hoekstra}
\affiliation{University of Groningen, KVI, Zernikelaan 25, 9747 AA, Groningen, The Netherlands}
\author{Hendrick L. Bethlem}
\affiliation{LaserLaB, Department of Physics and Astronomy, VU University Amsterdam,
De Boelelaan 1081, 1081 HV Amsterdam, The Netherlands} 

\date{\today}

\begin{abstract}
We have recently demonstrated static trapping of ammonia isotopologues in a decelerator that consists of a series of ring-shaped electrodes to which oscillating high voltages are applied [Quintero-P\'{e}rez~\emph{et al.}, \href{http://prl.aps.org/abstract/PRL/v110/i13/e133003}{Phys. Rev. Lett. \textbf{110}, 133003 (2013)}]. In this paper we provide further details about this traveling wave decelerator and present new experimental data that illustrate the control over molecules that it offers. We analyze the performance of our setup under different deceleration conditions and demonstrate phase-space manipulation of the trapped molecular sample.
\end{abstract}


\maketitle

\section{Introduction}
Cold molecules offer many exciting prospects in both chemistry and physics (for recent review papers see Refs.~\cite{Carr:NJP2009,Hogan2011,vandeMeerakker:ChemRev2012,Narevicius:ChemRev2012}). The great control that can be exerted over cold molecules allows the study and manipulation of collisions and chemical reactions~\cite{Ospelkaus2010,Kirste23112012}. The strong dipole-dipole interactions between cold molecules make them excellent systems for quantum simulation and computation~\cite{DeMille2002,Micheli2006,Andre2006}. With their rich structure, molecules can be useful systems for making precision tests of fundamental physics, such as the measurements of the electron EDM~\cite{Hudson:Nature2011,Campbell2013} and the search for time-variation of fundamental constants~\cite{Veldhoven2004,Hudson:PRL2006,Shelkovnikov2008,Truppe2013}. In particular this last application is of great interest to our group, and as a part of this research we are building a molecular fountain~\cite{Bethlem2008}.

A fountain requires great control over the molecules, in our case slowing them from around 300 m/s to rest, cooling them, and then launching them upwards at around 3 m/s. A tool that has been successfully used to exert control over molecules is the Stark decelerator~\cite{Bethlem1999,Bethlem:PRA2002,vandeMeerakker:ChemRev2012}. While excellent at removing kinetic energy, a Stark decelerator becomes lossy for low speeds ($<$100\,m/s) because of its reliance on creating an effective 3D potential well. When the characteristic wavelength of the longitudinal and transverse motion becomes comparable to the periodicity of the decelerator, the approximations used to derive this effective well no longer apply and both the number of molecules and phase-space density decline with speed~\cite{Sawyer:Thesis,Jansen:Thesis}. To avoid these losses at low speeds we use a traveling wave decelerator.

Based on the design of Osterwalder et al~\cite{Osterwalder2010,Meek2011}, this decelerator uses a sequence of ring-shaped electrodes to which a space- and time-varying voltage is applied. In this way the traveling wave decelerator creates a genuine rather than effective 3D trap that moves along the decelerator, co-propagating with the molecules. By reducing the speed of this co-moving trap the molecules are decelerated. The lowest speed reached by Osterwalder et al was 120 m/s, limited by the lowest frequency at which the voltage could be varied. The same decelerator was used to slow YbF molecules from a buffer gas source from 300 m/s to 276 m/s, limited by the deceleration that could be applied to these heavy molecules~\cite{Bulleid:PRA2012}. A 5\,m long traveling wave decelerator is currently under construction at the KVI in order to decelerate SrF molecules~\cite{Berg:EPJD2012}. 

We have recently been able to slow molecules to rest, and statically trap them inside the decelerator by using high voltage amplifiers that can sweep the high voltage from 15\,kHz down to DC. We use a conventional Stark decelerator to slow ammonia molecules from 300 m/s to around 100 m/s, removing 90\% of their initial kinetic energy over a length of only 50 cm. We then load the molecules into the traveling wave decelerator for the remaining deceleration to standstill~\cite{QuinteroPerez2013}.

In this paper we provide further details about the traveling wave decelerator, and present new experimental data that illustrate the control over molecules that is offered by this decelerator.

\section{Experimental setup}
\begin{figure}[tbh]
\centering
\includegraphics[width=1\columnwidth]{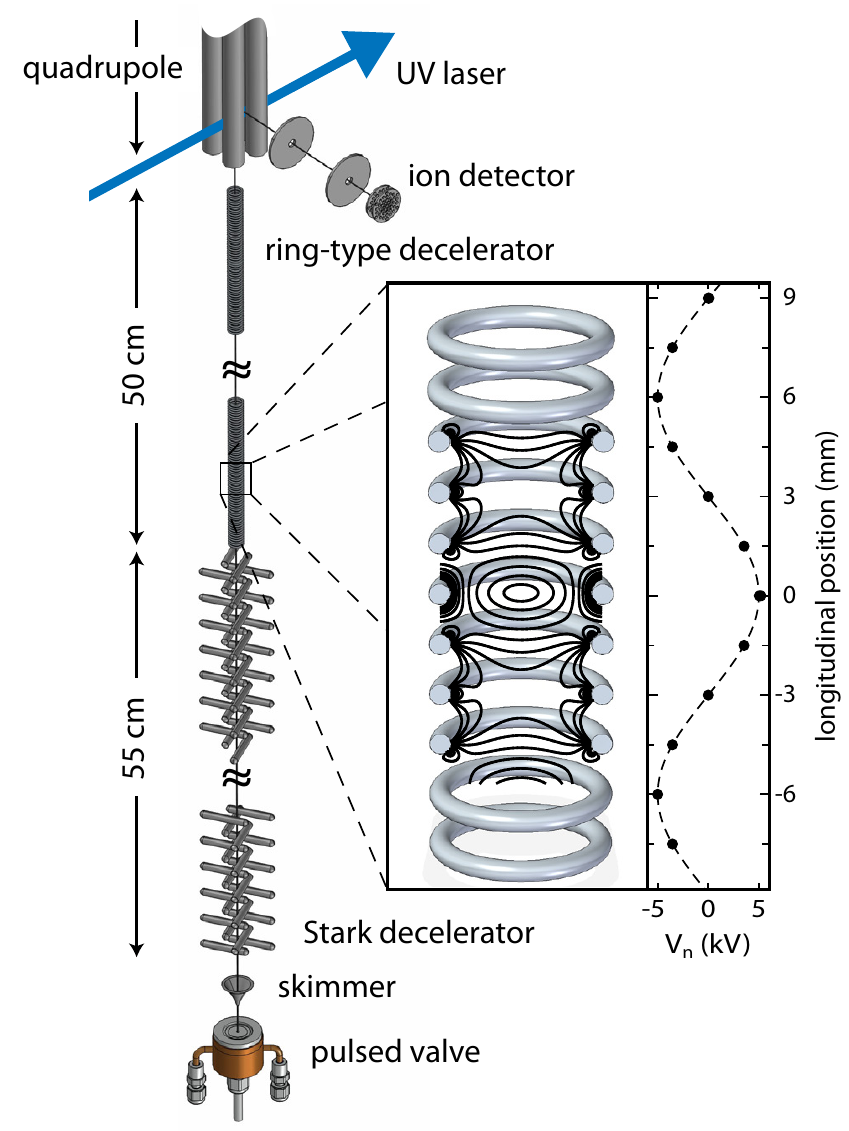}
\caption{(color online) Schematic view of the vertical molecular beam machine. The inset shows the electric field magnitude (in steps of 2.5\,kV/cm) inside the traveling wave decelerator calculated using {\sc Simion} \cite{Dahl:RevSciInstrum1990}. The rightmost panel shows the sinusoidal waveform (dashed line) from which the potential that is applied to each electrode (solid circle) is sampled.
\label{fig:setup_ring}}
\end{figure}

Figure~\ref{fig:setup_ring} shows a schematic view of our vertical molecular beam machine. In brief, a pulsed ($\sim$100\,$\mu$s) ammonia beam is released into vacuum from a solenoid valve (General Valve series 99) at a 10\,Hz repetition rate. By cooling the valve housing to typically -50$^{\circ}$\,C and seeding the ammonia molecules in xenon, the mean velocity of the beam is lowered to 300\,m/s. The ammonia beam is decelerated using a 101-stage Stark decelerator. Adjacent stages are 5.5\,mm apart. Each deceleration stage is formed by two parallel 3\,mm diameter cylindrical rods, spaced 2\,mm apart. The two opposite rods are switched to $+10$ and $-10$\,kV by four independent HV switches that are triggered by a programmable delay generator. 
A traveling wave decelerator is mounted 24~mm above the last electrode pair. This decelerator consists of 336 ring electrodes, each of which is attached to one of eight 8~mm diameter stainless steel rods, resulting in a periodic array in which every ninth ring electrode is attached to the same rod. The rods are placed on a 26\,mm diameter circle, forming a regular octagon. Each rod is mounted by two ceramic posts that are attached to the octagon via an adjustable aluminum bar that allows for fine-tuning of the alignment. The ring-shaped electrodes are made by bending 0.6~mm thick tantalum wire into the shape of a tennis racket with an inner diameter of 4~mm. Consecutive rings are separated by 1.5~mm (center to center) resulting in a periodic length of $L=12$\,mm. This combination of parameters is identical to the design of Osterwalder and co-workers~\cite{Osterwalder2010,Meek2011}. 

The voltages applied to the eight support bars are generated by amplifying the output of an arbitrary wave generator (Wuntronic DA8150) using eight fast HV-amplifiers (Trek 5/80) up to $\pm$5~kV. A 50\,cm long quadrupole is mounted 20\,mm above the traveling wave decelerator. The quadrupole can be used for focusing slow molecules and to provide an extraction field for a Wiley-McLaren type mass spectrometer setup. In the experiments described here, slow molecules are not focused and the quadrupole is used only to create the extraction field. The molecular beam overlaps with the focus of an UV laser 40\,mm behind the last ring electrode of the decelerator to ionize the ammonia molecules. The nascent ions are counted by an ion detector. The chamber that houses the two decelerators and quadrupole guide is differentially pumped and kept at a pressure below $3\times10^{-8}$\,mbar when the pulsed valve is operating. 

\section{Theory\label{sec:theory}}
A detailed discussion of the operation principles of the traveling wave decelerator can be found elsewhere~\cite{Osterwalder2010,Meek2011,vandeMeerakker:ChemRev2012}. In this section we will summarize those results that are relevant for the current paper. \\

In order to describe the operation of a traveling wave decelerator, it is useful to consider the electric field inside an infinitely long hollow conducting cylinder to which a voltage is applied that is periodic in $z$. As shown in Ref.~\cite{vandeMeerakker:ChemRev2012} the electric potential on the beam axis will in this case follow the potential applied to the cylinder, but it is reduced by a factor that depends on the radius of the cylinder and the periodic length of the applied waveform. The electric field magnitude in the longitudinal direction is given by a fully-rectified sine wave, resulting in two minima per period at the positions where the electric potential at the cylinder is maximal. At the position of these minima, the  electric field magnitude in the radial direction is given by a first-order modified Bessel function of the first kind. Close to the minima the electric field increases linearly in both directions with the field gradient in the longitudinal direction being twice as large as the gradient in the radial direction. These minima will act as true 3D traps for weak-field seeking molecules. By modulating the waveform that is applied to the cylinder in time, the traps can be moved along the decelerator, while keeping a constant shape and depth.

In the actual implementation of the traveling wave decelerator, 8 ring-shaped electrodes are used to sample the infinite cylinder. The inset of Fig.~\ref{fig:setup_ring} shows the electric field magnitude inside the traveling wave decelerator at a given time. 
The voltages applied to successive ring electrodes follow a sinusoidal pattern in $z$ shown on the right-hand side of the figure. In order to move the traps, the sinusoidal waveform is modulated in time. As a consequence of using a finite number of electrodes, the trapping potential no longer maintains a constant shape and depth while it is moved. In the chosen geometry, the electric field gradients in the bottom of the well, as well as the trap depth in the longitudinal direction are nearly independent of the position of the trap minimum. The trap depth in the transverse direction, however, is 40\% deeper when the trap minimum is located in the plane of a ring compared to the situation when the trap minimum is located directly between two rings.  

\begin{figure*}[tbh]
\centering
\includegraphics[width=0.7\textwidth]{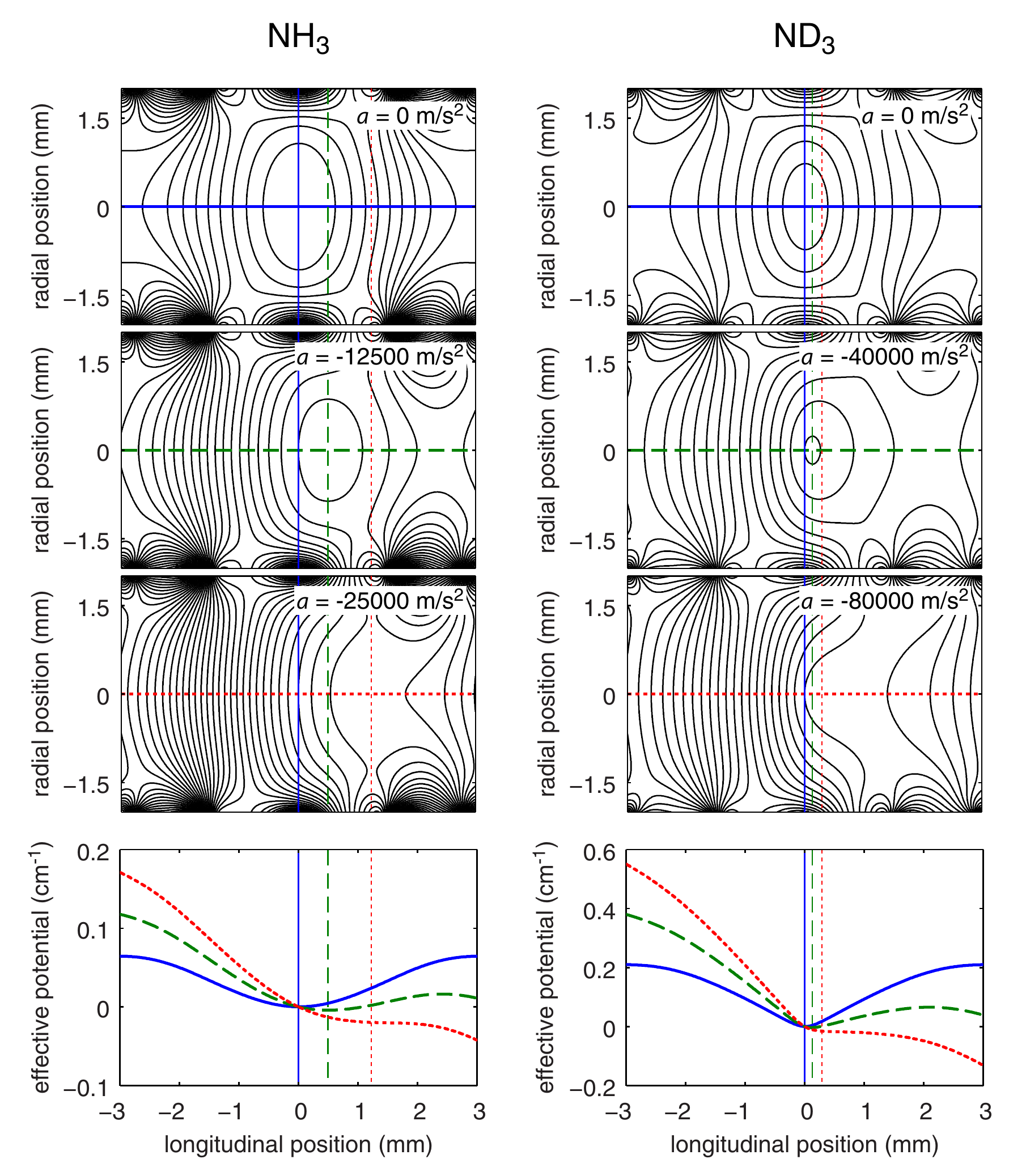}
\caption{(color online) Effective potentials for \ce{NH3} (left) and \ce{ND3} (right) for various accelerations and a waveform amplitude of 5\,kV. Contour lines are separated by 10\,mK for \ce{NH3} and 40\,mK for \ce{ND3}. A constant acceleration produces a pseudoforce that distorts the Stark potential and reduces the trap volume and depth. Above a certain threshold the potential no longer contains a minimum and no molecules can be trapped. The lower panels show the longitudinal dependence of the effective trapping potential along the beam axis for the different accelerations (relative to $a=0$\,m/s). Note the difference in the position of the trap center between \ce{NH3} and \ce{ND3} for $a\neq 0$\,m/s$^2$.
\label{fig:effectivepotential}}
\end{figure*}

The potential $V_n$ applied to the $n$th electrode in the traveling wave decelerator can be expressed as~\cite{Meek2011}

\begin{equation}
V_n(t)=V_0\sin\left (-\phi(t)+\frac{2\pi n}{8}\right ),
\label{eq:fields}
\end{equation}

\noindent
where $V_0$ is the amplitude of the sine-modulated potential and $\phi$ is a time-dependent phase offset that governs the motion of the traps. For the electric field configuration that is shown in Fig.~\ref{fig:setup_ring}, the phase offset has a value of $\phi=0 \pmod{{\pi}/{4}}$. The angular frequency of the wave is given by the time derivative of $\phi(t)$; $(d\phi/dt)=2\pi f(t)$, where $f(t)$ is the frequency in Hz. Since one oscillation of the waveform moves the trap over one period, the velocity of the trap is given by $v_z(t)=f(t) L$. Integrating the angular frequency with respect to time results in the following expression for the phase

\begin{equation}
\phi(t)=\frac{2\pi}{L}\int_0^t\! v_z(\tau)\,d\tau,\quad\text{where }v_z(t)=\int_0^t\!a(\tau)\,d\tau. 
\end{equation}

\noindent
A linear increase of $\phi$ with time results in a trap that moves with a constant positive velocity along the $z$ axis. Acceleration or deceleration of the trap can be achieved by applying a chirp to the frequency. 

The acceleration or deceleration of the trap changes the effective longitudinal potential experienced by the molecules. In order to account for this pseudoforce, an additional term of the form $W_{\text{acc}}=maz$, with $m$ the mass of the molecule and $a$ the acceleration along the $z$ axis, is added to the potential.  In Fig.~\ref{fig:effectivepotential}, the resulting potentials are shown for different waveforms corresponding to different accelerations as indicated. The panels on the left- and right-hand side show the potential experienced by \ce{NH3} and \ce{ND3} molecules, respectively. The lower panels show the effective potential along the $z$-axis. 
The inversion splitting in NH$_{3}$ is 23.8\,GHz, while it is only 1.6\,GHz in ND$_{3}$. As a result, the Stark effect in NH$_{3}$ is quadratic up to electric fields of 20\,kV/cm and the effective potential for NH$_{3}$ molecules is almost perfectly harmonic. The Stark effect in ND$_{3}$, on the other hand, becomes linear at much smaller fields and the effective potential for ND$_{3}$ molecules is harmonic close to the trap center only. 
Note that in the figure the accelerations applied to \ce{ND3} are about 3 times larger than those applied to \ce{NH3}. As seen from the figure, acceleration of the trap reduces its phase-space acceptance and thus decreases the number of molecules that can be confined. Above a certain threshold value, the potential no longer contains a minimum and no molecules can be trapped at all. In addition to reducing the trap depth, acceleration of the trap results in a shift of the effective field minimum. This shift needs to be taken into account when the acceleration is not constant but a function of time, for instance, when the molecules are decelerated to a standstill and subsequently trapped at a certain position in the decelerator. We do this by rapidly ($<20$\,$\mu$s) sweeping the phase of the waveform. We will refer to this sweep as a ``phase jump''.

\section{Results}
In this section we present data that demonstrate the versatility of the traveling wave decelerator and analyze the performance of the setup under different conditions. Results for guiding, deceleration and trapping of \ce{NH3} and \ce{ND3} molecules are presented in Section~\ref{sec:guidingdectrap}. Section~\ref{sec:phasejumps} discusses the use of the phase jump for \ce{NH3} and \ce{ND3}.

\begin{figure}[bth!]
\centering
\includegraphics[width=1\columnwidth]{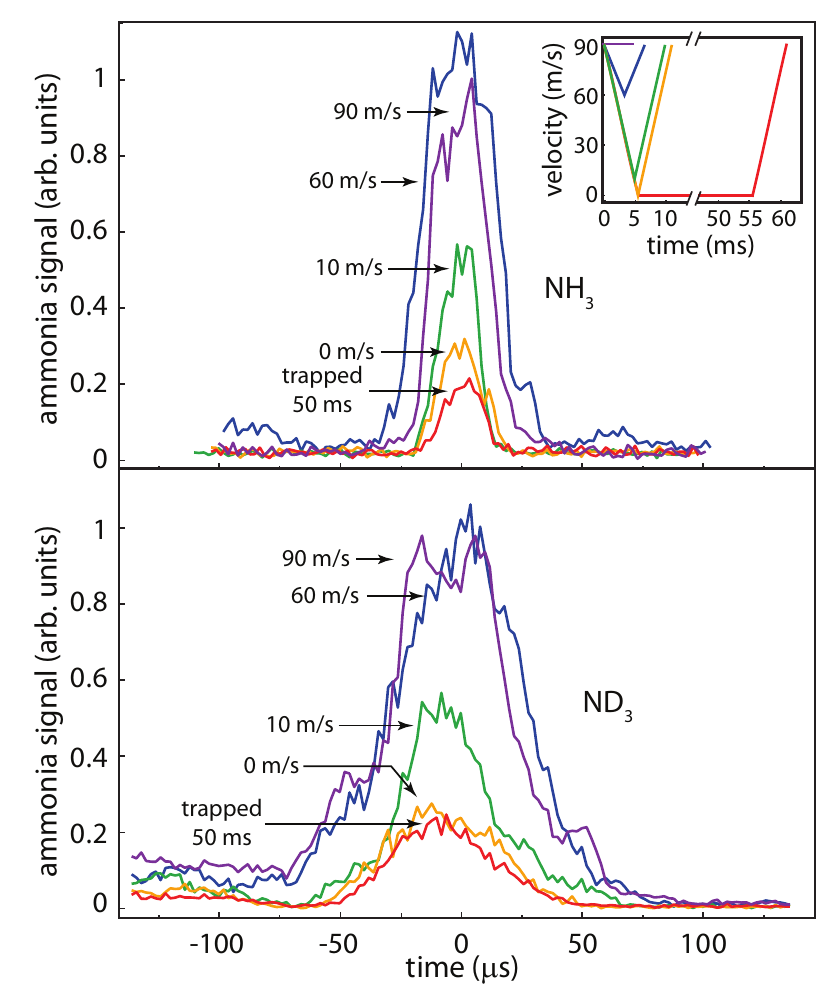}
\caption{(color online) Measured time-of-flight profiles for \ce{NH3} (upper panel) and \ce{ND3} (lower panel) molecules guided at 90\,m/s (violet curve), decelerated to 60, 30 and 0\,m/s (blue, green, and yellow curves respectively) and trapped for 50\,ms (red curve) before being accelerated back to 90\,m/s and detected. Each data point is the average of 90 laser shots. The time of flight traces have been centered around the expected arrival time. The inset shows the velocity of the traveling wave potential as a function of time for the different recorded TOF profiles.
\label{fig:tofs}}
\end{figure}

\subsection{Guiding, deceleration and trapping\label{sec:guidingdectrap}}
In Fig.~\ref{fig:tofs}, time-of-flight (TOF) profiles of \ce{NH3} molecules (upper panel) and \ce{ND3} molecules (lower panel) are shown that were obtained by applying different waveforms to the traveling wave decelerator, as shown in the inset. Note that the horizontal axis is shifted in such a way that the origin corresponds to the expected arrival time of the molecules. For each measurement, a packet of molecules is decelerated to 90\,m/s using the conventional Stark decelerator and subsequently coupled into the traveling wave decelerator. The violet curves in Fig.~\ref{fig:tofs} are TOF profiles for packets of \ce{NH3} and \ce{ND3} molecules that are guided through the traveling wave decelerator at the injection speed of 90\,m/s by applying a waveform with a constant frequency of 7.5\,kHz to the array of electrodes. The width of the TOF profile mainly reflects the velocity spread of the guided molecules. For \ce{ND3} the moving trap is deeper than for \ce{NH3}, and its TOF profile is accordingly wider. Wings are observed at earlier and later arrival times, which are attributed to molecules that are trapped in the electric-field minima that are located $12$\,mm above or below the synchronous one.  

By adjusting the waveform that is applied to the traveling wave decelerator, the velocity of the molecules can be manipulated almost at will. The blue, green, and yellow traces in Fig.~\ref{fig:tofs} are time-of-flight profiles for \ce{NH3} and \ce{ND3} molecules that are decelerated from 90\,m/s to 60, 10 and 0\,m/s and immediately reaccelerated to 90\,m/s. This corresponds to chirping the applied waveform from 7.5\,kHz to 5.0, 0.8, and 0\,kHz and results in accelerations of $\pm$9.2, $\pm$16.4, and $\pm 16.6\times 10^{3}$\,m/s$^2$, respectively.

The red curves in Fig.~\ref{fig:tofs} show TOF profiles that are obtained under almost identical conditions as the yellow curves. However, after the velocity of the applied waveform is decreased to 0\,m/s, the voltages are kept at a constant values for 50\,ms before the velocity is increased to its original value of 90\,m/s. It can be seen in the figure that the observed TOFs are indeed almost identical to the ones recorded when the velocities are immediately returned to their original value. This measurement demonstrates that molecules can be trapped in the laboratory frame without further losses. 

The observed decrease in signal for NH$_{3}$ and ND$_{3}$ at higher accelerations is greater than expected from simulations. We attribute the loss mainly to mechanical misalignments that lead to parametric amplification of the motion of the trapped molecules at low velocities. On inspection, it was noticed that one of the suspension bars was slightly displaced from its original position, which must have happened when the decelerator was placed in the vacuum chamber. Another loss mechanism comes from the fact that the phase space distribution of the packet exiting the conventional Stark decelerator is not perfectly matched to the acceptance of the traveling wave decelerator, and the alignment of the axis of the traveling wave decelerator to the axis of the Stark decelerator is not perfect. As a result, the trapped packet as a whole will perform a (damped) breathing and sloshing motion. These oscillations explain why the observed \ce{NH3} TOF profile for deceleration to 60\,m/s is wider and more intense than the TOF profile for guided molecules. Neither loss mechanism is fundamental, and we believe they can be eliminated in future work. 

\begin{figure}[bth!]
\centering
\includegraphics[width=1\columnwidth]{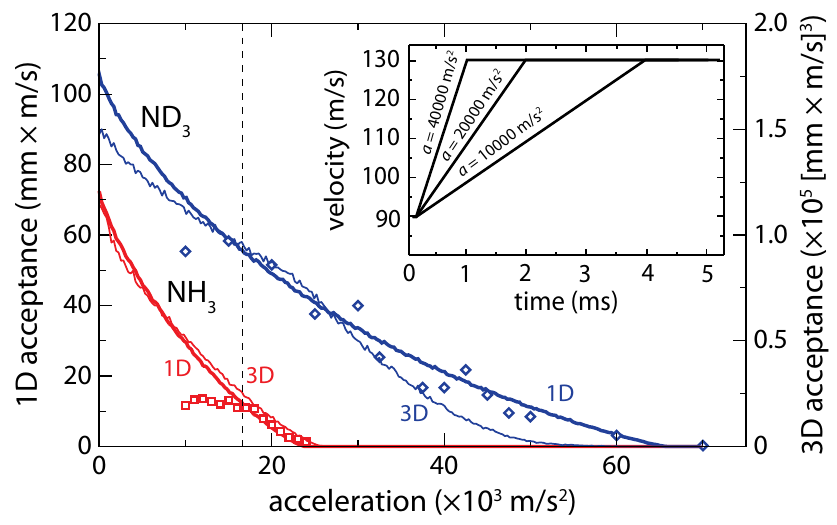}
\caption{(color online) Integrated time-of-flight distributions for \ce{NH3} (squares) and \ce{ND3} (diamonds) shown as a function of acceleration. The bold lines correspond to simulations of the 1D  acceptance of the traveling wave decelerator, shown on the vertical axis on the left-hand side, while the thin lines correspond to simulations of the 3D acceptance of the traveling wave, shown on the vertical axis on the right-hand side. The \ce{NH3} and \ce{ND3} signals are scaled (by a different factor) to match the simulation. The inset displays the velocity of the traveling wave potential for three typical accelerations used in the experiment. 
\label{fig:accvssignal}}
\end{figure}

As discussed in Sec.~\ref{sec:theory}, the acceptance of the traveling wave decelerator depends on the chirp that is applied to the waveform. This is illustrated by the measurements that are shown in Fig.~\ref{fig:accvssignal}. In this figure, the time-integrated signal of \ce{NH3} (squares) and \ce{ND3} (diamonds) molecules is plotted as a function of the applied acceleration. In these measurements, ammonia molecules are guided at 90\,m/s for 5\,mm before being accelerated to 130\,m/s and guided for the remaining length of the decelerator.  

For comparison, the bold and thin solid lines shown in Fig.~\ref{fig:accvssignal} correspond to simulations of the 1D and 3D acceptance of the traveling wave decelerator, respectively. The calculated acceptances are based on an average (i.e. static in the moving frame of the trap) potential well, rather than the true potential. The numbers on the right axis refer to the 3D acceptance, while the numbers on the left-hand axis refer to the 1D acceptance. The fact that the 1D and 3D simulations are almost the same, apart from a scaling factor, illustrates that the transverse acceptance is independent of the acceleration, i.e., the transverse motion of the molecules in the traveling wave decelerator is largely decoupled from the longitudinal motion. The measurements for \ce{ND3} and \ce{NH3} have been scaled to match the simulations. As observed, at high acceleration the simulations predict the measurements quite well. At lower accelerations, however, the measurements are seen to reach a constant value which is consistent with the expected longitudinal acceptance of the conventional Stark decelerator at the used phase angle of 65 degrees~\cite{Bethlem:PRA2002}. At these accelerations the longitudinal acceptance of the traveling wave decelerator is larger than the longitudinal emittance of the packet exiting the Stark decelerator; i.e., all molecules exiting the conventional Stark decelerator are trapped in the traveling wave decelerator. Consequently, decelerations below roughly $18\times 10^3$ m/s$^{2}$ do not reduce the detected signal of \ce{NH3} and \ce{ND3} molecules. The dashed vertical line also shown in Fig.~\ref{fig:accvssignal} indicates the maximum acceleration that was used in the measurements shown in Fig.~\ref{fig:tofs}. Note that, as in our measurements we do not determine an absolute value for the phase space acceptance of the decelerator, the similarity between the measurements and simulations should be considered as a consistency check only.  

\begin{figure}[bth!]
\centering
\includegraphics[width=1\columnwidth]{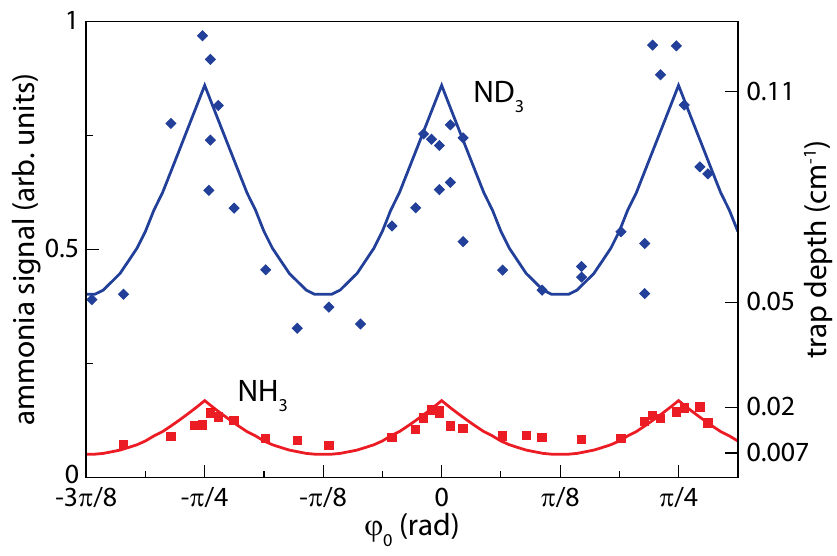}
\caption{(color online) \ce{NH3} (squares) and \ce{ND3} (diamonds) signal as a function of the position in the decelerator (expressed as a phase, $\phi_0$) at which the molecules were trapped. When $\phi_0=0\pmod{{\pi}/{4}}$, the center of the trap is located in the plane of a ring electrode while for $\phi_0=\pi/8 \pmod{{\pi}/{4}}$ the trap minimum is located in the plane between two ring electrodes. The solid curves represent the trap depth for both species and are specified on the right axis. The \ce{NH3} and \ce{ND3} signals are scaled (by different factors) to match the trap depth. 
\label{fig:positionscan}}
\end{figure}

The position in the array of ring-shaped electrodes at which the molecules are brought to a standstill and statically trapped, is determined by the waveform and can be chosen at will. Figure~\ref{fig:positionscan} shows the \ce{NH3} (squares) and \ce{ND3} (diamonds) signal as a function of this position. The signal oscillates with a periodicity that corresponds to the spacing between adjacent rings; 1.5\,mm equivalent to a phase difference of $\pi/4$. The signal is largest when the ammonia molecules are trapped in the plane of an electrode, and smallest when the trap center is located between two electrodes. The observed modulation is related to the variation in the radial confinement discussed in Sec.~\ref{sec:theory}. For comparison, the solid lines in Fig.~\ref{fig:positionscan} show the trap depth for \ce{NH3} and \ce{ND3} in cm$^{-1}$ (1~cm$^{-1}$ corresponds to 1.4\,K). The \ce{NH3} and \ce{ND3} signals are scaled (by different factors) to match the trap depth.

\subsection{Phase jumps\label{sec:phasejumps}}

\begin{figure}[tbh]
\centering
\includegraphics[width=1\columnwidth]{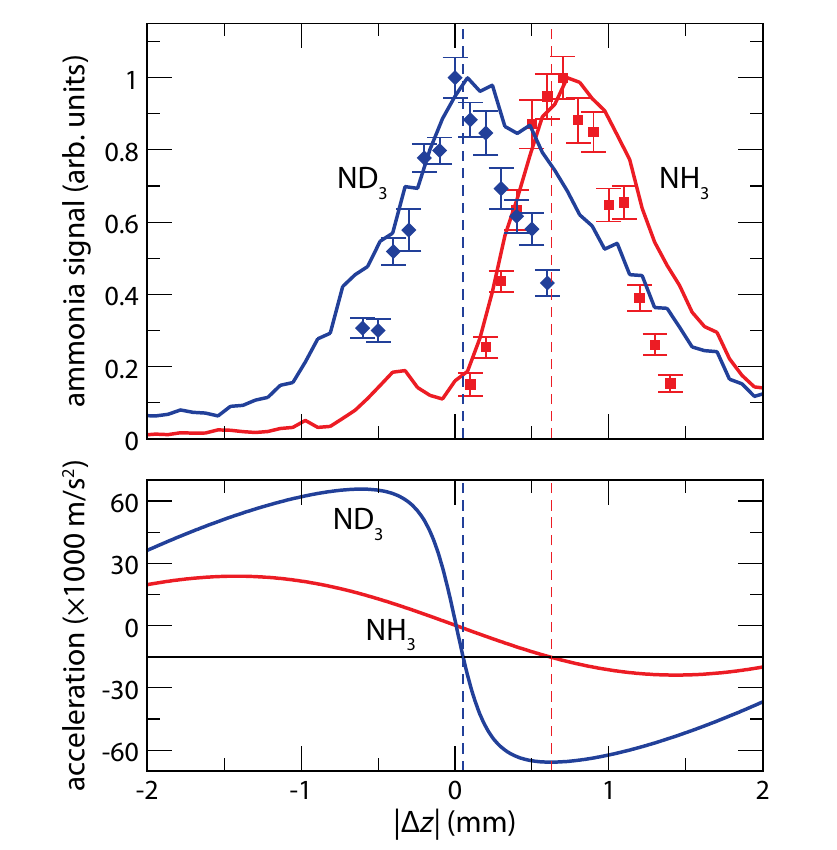}
\caption{(color online) The effect of phase jumps for an acceleration of -15\,000\,m/s$^2$ are illustrated in the upper panel for \ce{NH3} (squares) and \ce{ND3} (diamonds). Solid curves show the result of a 1D numerical simulation. The lower panel shows the longitudinal acceleration along the beam axis for \ce{NH3} and \ce{ND3} molecules using a waveform amplitude of 5\,kV. The dashed vertical lines indicate the phase jumps that are derived from the acceleration curves in the lower panel.
\label{fig:phasejump}}
\end{figure}

As explained in Sec.~\ref{sec:theory}, whenever the acceleration is changed, we rapidly sweep the phase of the waveform in order to correct for the resulting shift of the effective potential minimum. We refer to this sweep as a ``phase jump''. In Fig.~\ref{fig:phasejump} we present measurements of molecules that are decelerated from 70 to 0\,m/s using an acceleration of -15000\,m/s$^2$, trapped for 30\,ms, and subsequently accelerated back to 70\,m/s with the opposite acceleration and detected. The upper panel of Fig.~\ref{fig:phasejump} shows the \ce{NH3} (squares) and \ce{ND3} (diamonds) signal as a function of the (magnitude of the) phase jumps that are applied whenever the acceleration is changed. As expected, the required phase jump for optimal signal is much larger for \ce{NH3} than for \ce{ND3}. In fact, when no phase jump is applied almost no signal is observed for \ce{NH3}, while for \ce{ND3} the signal hardly changes. The solid curves in the upper panel of Fig.~\ref{fig:phasejump} show the result of 1D numerical simulations. Although the simulations seem to slightly overestimate the width of the observed peaks and predict slightly greater phase jumps, the maximum and overall shape match the experimental data fairly well. The discrepancies might be caused by the fact that the amplifiers are not able to follow the rapid sweep of the phase perfectly. 

The lower panel of Fig.~\ref{fig:phasejump} shows the longitudinal acceleration as a function of the position along the beam axis for \ce{NH3} and \ce{ND3} molecules using a waveform amplitude of 5\,kV. The electric field increases linearly away from the center of the trap resulting in a harmonic but shallow potential for \ce{NH3} and a tight but very anharmonic potential for \ce{ND3}. The horizontal line shows the acceleration that is used in the experiment. The crossings of this line with the acceleration curves, indicated by the dashed vertical lines, correspond to the positions of the minima of the effective potential.

\subsection{Adiabatic cooling\label{sec:cooling}}
\begin{figure}[tb!]
\centering
\includegraphics[width=1\columnwidth]{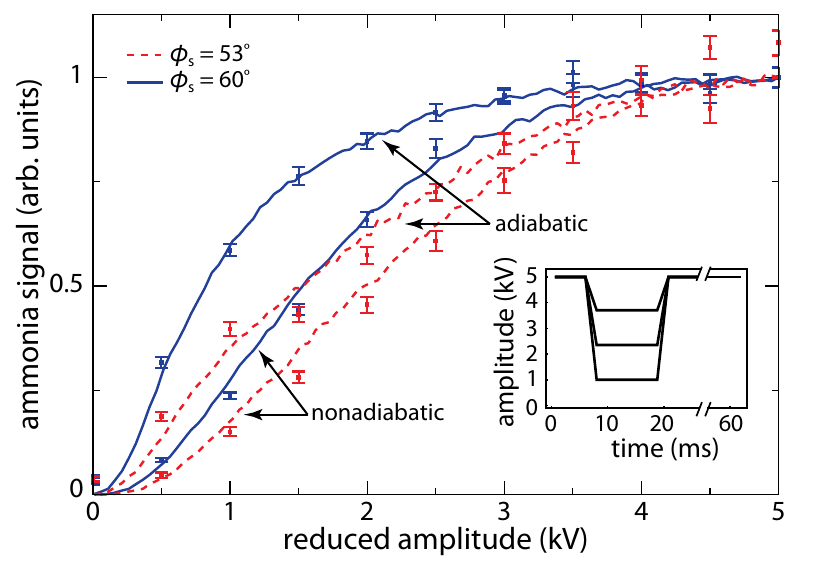}
\caption{(color online) Signal of trapped \ce{ND3} molecules decelerated using a phase angle of 60 and 53 degrees as a function of the amplitude of the waveform when, after deceleration, the trap is slowly (labeled as adiabatic) or abruptly (labeled as nonadiabatic) reduced.  The smooth curves also shown in the figure are simulations that assume an initial temperature of 30 (solid line) or 100\,mK (dashed line) for the ensemble of trapped \ce{ND3} molecules. The inset shows the amplitude of a number of typical waveforms as a function of time.
\label{fig:cooling}}
\end{figure}

As the voltages applied to the decelerator are generated by amplifying the output of an arbitrary wave generator using fast HV-amplifiers, we can change the depth (and shape) of the trap at will. This is illustrated by the measurements shown in Fig.~\ref{fig:cooling}. In these measurements, \ce{ND3} molecules were decelerated using the conventional Stark decelerator at phase angles of 60 and 53 degrees and coupled into the traveling wave decelerator. The traveling wave decelerator is subsequently used to slow the \ce{ND3} molecules to a standstill, trap them for a period of over 50\,ms, accelerate them back to their initial velocity and launch them into the detection region. While the molecules are trapped, the voltages applied to the decelerator are ramped down in 2\,ms, kept at a lower value for 10\,ms and then ramped up again to 5\,kV. Figure~\ref{fig:cooling} shows the remaining \ce{ND3} signal as a function of the reduced amplitude of the waveform for the two phase angles as indicated. Typical waveform amplitudes as a function of time are shown in the inset of Fig.~\ref{fig:cooling}. Lowering the voltages of the trap has two effects: (i) the trap frequency is lowered, adiabatically cooling the molecules; (ii) the trap depth is reduced, allowing the hottest molecules to escape the trap. The observed signal at a phase angle of 53 degrees is seen to drop more rapidly than that at a phase angle of 60 degrees. This is expected, as at lower phase angles a larger and hotter packet is loaded into the traveling wave decelerator. The hotter ensemble of \ce{ND3} molecules initially fills the trap almost completely while the colder molecules occupy only a fraction of the trap. Consequently, the number of molecules in the hotter ensemble is reduced even when the voltages are lowered by a small amount. The solid and dashed lines also shown in Fig.~\ref{fig:cooling} result from a simulation that assumes an (initial) temperature of 30\,mK or 100\,mK, respectively. Note that we use temperature here only as a convenient means to describe the distribution; the densities are too low to have thermalization on the timescales of the experiment. For comparison, Fig.~\ref{fig:cooling} also shows measurements where the trap voltages are abruptly (10\,$\mu$s) lowered. In this case no adiabatic cooling occurs and as a result the signal decreases more rapidly than when the voltages are ramped to lower voltages more slowly. The solid (30\,mK) and dashed (100\,mK) lines labeled as nonadiabatic show the results of a numerical simulation when the voltages are reduced abruptly. Similar measurements have been performed for \ce{NH3} (not shown).

\begin{figure}[tbh]
\centering
\includegraphics[width=1\columnwidth]{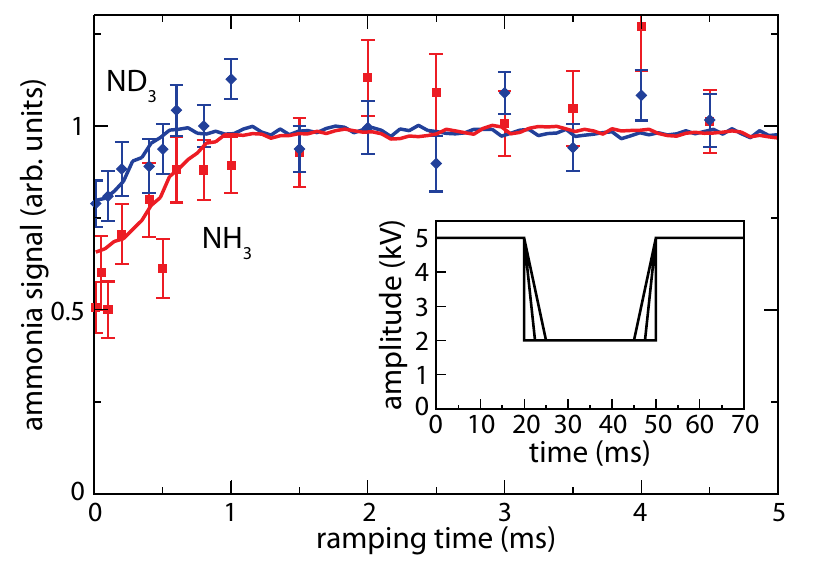}
\caption{(color online) Signal due to \ce{NH3} (squares) and \ce{ND3} (diamonds) molecules that remain trapped at 2\,kV as a function of the time used for ramping the voltages. The solid lines show the result of a 3D numerical simulation that assumes a temperature of 30 or 100\,mK for \ce{NH3} and \ce{ND3}, respectively. In the inset, typical waveform amplitudes are shown as a function of time.
\label{fig:tramp}}
\end{figure}

In Fig.~\ref{fig:tramp}, the number of molecules that remain trapped at 2\,kV is shown as a function of the time used for lowering and increasing the voltages. The solid lines result from a 3D numerical simulation. The measurements confirm that the \ce{NH3} and \ce{ND3} molecules follow the trap adiabatically when the ramping times are longer than 1\,ms. The time required for the molecules to follow the change in potential adiabatically is greater for \ce{NH3} than it is for \ce{ND3} due to the lower trap frequencies experienced by the former. Similar measurements have been performed on \ce{CH3F} molecules in a microstructured trap by Englert \emph{et. al}~\cite{Englert2011}. 

\section{Conclusions}
In this paper we have demonstrated deceleration and trapping of \ce{NH3} and \ce{ND3} molecules in a traveling wave decelerator. The observed trapping times are limited to 100\,ms by the current repetition rate of the experiment (10\,Hz). The deceleration of a supersonic beam of ammonia molecules is performed in two steps; in the first step, the molecules are decelerated from 300 to 100\,m/s using a conventional Stark decelerator, while in the second step, the traveling wave decelerator is used to remove the remaining kinetic energy from the molecules. The advantages of such a system are that the requirements imposed on the electronics of the traveling wave decelerator remain rather low, and that the combined length of both decelerators is only slightly above 1\,m. 

The motion of the molecules in the traveling wave decelerator is controlled completely by computer generated waveforms. As an example of the possibilities offered by this control, we have adiabatically cooled the trapped molecules by lowering the amplitude of the waveforms. In previous work, we demonstrated that the motion  of the molecules in the trap can be excited by applying an small  oscillatory force~\cite{QuinteroPerez2013}. The ability to control the voltages applied to the electrodes also offers the possibility to tailor the shape of the trap -- for instance, changing it into a more box-like potential. This may prove useful for collisional and spectroscopic studies, as well as for the implementation of schemes to further cool the molecules such as sisyphus cooling~\cite{Zeppenfeld:Nature2012} and evaporative cooling~\cite{Stuhl:Nature2012}. 

\begin{acknowledgments}
This research has been supported by NWO via a VIDI-grant, by the ERC via a Starting Grant and by the FOM-program `Broken Mirrors \& Drifting Constants'. We acknowledge the expert technical assistance of Rob Kortekaas, Jacques Bouma, Joost Buijs, Leo Huisman and Imko Smid. We thank Andreas Osterwalder and Gerard Meijer for helpful discussions and Wim Ubachs for his continuing interest and support.
\end{acknowledgments}

\end{document}